# String Theory

–

# Nomological Unification

# and the Epicycles of the Quantum Field Theory Paradigm[1]

**Reiner Hedrich**

Institut für Philosophie und Politikwissenschaft
Technische Universität Dortmund
Emil-Figge-Strasse 50
44227 Dortmund
Germany
reiner.hedrich@udo.edu

Zentrum für Philosophie und Grundlagen der Wissenschaft
Justus-Liebig-Universität Giessen
Otto-Behaghel-Strasse 10 C II
35394 Giessen
Germany
Reiner.Hedrich@phil.uni-giessen.de

**Abstract:**
*String Theory* is the result of the conjunction of three conceptually independent elements: (i) the metaphysical idea of a nomological unity of the forces, (ii) the model-theoretical paradigm of Quantum Field Theory, and (iii) the conflict resulting from classical gravity in a quantum world – the motivational starting point of the search for a theory of Quantum Gravity. *String Theory* is sometimes assumed to solve this conflict: by means of an application of the (only slightly extended) model-theoretical apparatus of (perturbative) Quantum Field Theory, interpreting gravity as the result of an exchange of gravitons, taken here to be dynamical states of the string. But, *String Theory* does not really solve the conflict. Rather it exemplifies the inadequacy of the apparatus of Quantum Field Theory in the context of Quantum Gravity: After several decades of development it still exists only in an essentially perturbative formulation (with minor non-perturbative extensions and vague ideas with regard to a possible non-perturbative formulation). And, due to its quantum field theoretical heritage, it is conceptually incompatible with central implications of General Relativity, especially those resulting from the general relativistic relation between gravity and spacetime. All known formulations of *String The-*

---

[1] This paper is based on research generously supported by the *Fritz-Thyssen-Stiftung für Wissenschaftsförderung* under the project *Raumzeitkonzeptionen in der Quantengravitation* (2007-2009) as well as by the *Deutsche Forschungsgemeinschaft* under the project FA 261/7-1 *Vereinheitlichung in der Physik durch den Superstring-Ansatz: Wissenschaftstheoretische und naturphilosophische Analyse* (2004-2006). Preliminary research was carried out during my stay (2002) as a Visiting Fellow at the *Center for Philosophy of Science* of the *University of Pittsburgh*. Thanks also to Brigitte Falkenburg!

*ory* are background-dependent. And no physical motivation is given for this conceptual incompatibility.

On the other hand, although *String Theory* identifies all gauge bosons as string states, it was not even possible to reproduce the Standard Model. Instead, *String Theory* led to a multitude of internal problems – and to the plethora of low-energy scenarios with different nomologies and symmetries, known as the *String Landscape*. All attempts to find a dynamically motivated selection principle remained without success, leaving *String Theory* without any predictive power. The nomological unification of the fundamental forces, including gravity, is only achieved in a purely formal way within the model-theoretical paradigm of Quantum Field Theory – by means of physically unmotivated epicycles like higher dimensionality, Calabi-Yau spaces, branes, etc.

Finally, the possibility remains that some of the central (implicit) assumptions of *String Theory* are physically wrong. On the one hand, the idea of a nomological unity of the forces could be simply wrong. On the other hand, even if a nomological unity of all *fundamental* forces should be realized in nature, the possibility remains that gravity is not a fundamental force, but a residual, emergent and possibly intrinsically classical phenomenon, resulting from a quantum substrate without any gravitational degrees of freedom.

Du temps perdu a la recherche ...

# 1. Quantum Gravity – The Motivations

The mutual conceptual incompatibility between General Relativity on the one hand and Quantum Mechanics and Quantum Field Theory on the other hand can be seen as the most essential motivation for the development of a theory of Quantum Gravity.[2]

One aspect of this incompatibility consists in the conflict resulting from classical gravity in a quantum world. General Relativity, today our best theory of gravity as well as of spacetime, treats the gravitational field as a classical dynamical field, represented by the (pseudo-) Riemannian metric of spacetime.[3] But, according to Quantum Mechanics, dynamical fields have quantum properties. So, if Quantum Mechanics is taken to be universally valid, it seems reasonable to assume the necessity of a (direct or indirect) quantization of the gravitational field. – An additional motivation for the quantization of gravity comes from rather conclusive arguments against semi-classical modifica-tions of the Einstein field equations, i.e. a formalism treating gravity classically and everything else quantum mechanically.[4]

---

[2] Under which conditions this conceptual incompatibility has to be seen as real or as only apparent, as well as what follows from each of these possibilities, will have to be discussed below; see section 2.2. – Further motivations for the development of a theory of Quantum Gravity come from specific physical problems, unsolved within the framework of the established theories and resulting at least partially from the fact that General Relativity predicts singularities: spacetime points for which it loses its validity. Here a theory of Quantum Gravity, by means of which we could get over the mutual incompatibility of General Relativity and Quantum Mechanics, seems to be inevitable. A successful, adequate theory of spacetime should be able to describe what happens in those cases. Such a theory should capture the presumed quantum properties of the gravitational field and of dynamical spacetime. Or it should be able to explain, how gravity and/or spacetime as possibly emergent, intrinsically classical phenomena with no quantum properties at all could be compatible with – and result from – a quantum world.

[3] All other fields as well as matter are also treated classically by General Relativity.

The other aspect of the conflict consists in the fact that neither Quantum Mechanics[5] nor Quantum Field Theory are compatible with essential conceptual insights of General Relativity. In General Relativity the gravitational field is represented by the metric of spacetime. Therefore, a quantization of the gravitational field would correspond to a quantization of the metric of spacetime. The quantum dynamics of the gravitational field would correspond to a dynamical quantum spacetime. But Quantum Field Theories presuppose a fixed, non-dynamical background space for the description of the dynamics of quantum fields. They are conceptually inadequate for a description of a dynamical quantum geometry. An attempt to find a quantum description of dynamical geometry by means of a theoretical approach that necessarily presupposes a background space with an already fixed metric will scarcely be successful. A quantum theory of the gravitational field can scarcely be a Quantum Field Theory, at least not one in the usual sense. – But it is not only the dynamical character of general relativistic spacetime that makes traditional background-dependent quantum theoretical approaches problematic. It is foremost the active diffeomorphism invariance of General Relativity that is fundamentally incompatible with any fixed background spacetime.[6]

# 2. Quantum Gravity – The Strategies

## 2.1. Looking for Quantum Properties of Gravity (and Spacetime)

If we take Quantum Mechanics seriously as our fundamental (and presumably universally valid) theory of the dynamics of matter and fields, it seems to be reasonable (at least at first sight) to assume that the gravitational field – like all other dynamical fields – should have quantum properties, not yet taken into account in the classical picture provided by General Relativity and to be identified by an adequate theory of Quantum Gravity.[7] But, what would be the most promising way to construct such a theory?

### 2.1.1. Direct Quantization of General Relativity

Taken into account the successful experiences with the implementation of all other fundamental interactions into a quantum mechanical description, leading to the (at least empirically) successful Standard Model of elementary particle physics, the most natural way to get to a theory of Quantum Gravity seems to be a more or less direct quantization of the gravitational field. Then, under the assumption that General Relativity can be seen as an adequate description of the classical aspects of gravity, a strategy which consists basically in a quantization of General Relativity can be seen as a natural avenue to a theory of Quantum Gravity. Considering that in General Relativity the gravitational field is represented by the metric of spacetime, i.e. that gravity is identical to properties of a dynamical geometry, a quantization of the gravitational field would then correspond to a quantiza-tion of the metric of spacetime. The quantum dynamics of the gravitational field would correspond to a dynamical quantum spacetime, a dynamical quantum geometry.[8]

---

[4] Cf. Kiefer (2004, 2005), Peres / Terno (2001), Terno (2006), Callender / Huggett (2001a, 2001b).

[5] One of the problems consists in the fact that Quantum Mechanics treats time as a global background parameter, not even as a physical observable represented by a quantum operator. In contrast, in General Relativity, time is a component of dynamical spacetime. It is dynamically involved in the interaction between matter/energy and the spacetime metric. It can be defined only locally and internally; there is no external global time parameter with physical significance.

[6] Cf. Earman (1986, 1989, 2002, 2006, 2006a), Earman / Norton (1987), Norton (1988, 1993, 2004).

[7] Much more clearly than this intuition, it is the already mentioned arguments against semi-classical theories of gravitation that exclude the possibility of a fundamental non-quantum gravitational interaction in a quantum world. – But these arguments are only valid, if gravity is a fundamental interaction.

But what would we have to expect with regard to the quantum properties of spacetime? – At least on first sight, our experiences with Quantum Mechanics taken again into account, we would probably suspect that the spacetime metric should be the expectation value of a quantum variable. On the quantum level, we would probably expect superpositions of spacetime metrics, uncertainties of spacetime, quantum fluctuations of the spacetime metric, of spacetime geometry, possibly even of the spacetime topology. And our experiences with Quantum Field Theories would possibly suggest some exchange boson for gravity: the 'graviton'. Quantum Gravity, one could think, should under these assumptions possibly be a theory describing the dynamics of gravitons exchanged between matter particles. – This is at least one of the ideas behind the *Covariant Quantization* program.

*Covariant Quantization*

The *Covariant Quantization*[9] of General Relativity consists in an attempt to construct a Quantum Field Theory of gravity, which means: a Quantum Field Theory of the metric field – analogously to Quantum Electrodynamics and its treatment of the electromagnetic field. But Quantum Field Theories in this orthodox sense need a background spacetime with a fixed metric for the definition of its operator fields. Consequently, *Covariant Quantization* uses a standard perturbation-theoretical approach, working with a fixed kinematical (usually Minkowski) background metric and a perturbation on this background to be treated quantum mechanically. This leads to a Quantum Field Theory of the fluctuations of the metric. The corresponding field quanta of gravity, the gravitons, are massless and have spin 2 – as a consequence of symmetry arguments and of the properties of classical gravity (long-range, exclusively attractive). They are assumed to represent the quantum properties of spacetime and to behave according to standard Feynman rules on a fixed background spacetime.

But the *Covariant Quantization* approach with its perturbation expansion of the fluctuations of the spacetime metric turns out to be non-renormalizable. This is not much of a surprise: Gravity couples to mass and, because of the mass-energy equivalence, to every form of energy.[10] Therefore the self-interaction contributions to gravity increase for decreasing distances or increasing energies. The contribution of virtual particles with increasing energies dominates the higher orders of the perturbation expansion. This leads to uncontrollable divergences of the expansion and to its non-renormalizability. No quantitative predictions can be achieved. This makes the theory irrelevant as a fundamental description of gravity and spacetime.

Obviously, it is not possible to get over the mutual conceptual incompatibility between General Relativity and Quantum Mechanics / Quantum Field Theory by simply amalgamating gravity and the quantum by means of the standard quantization procedures. The conceptual foundations of both are much too different. The *Covariant Quantization* approach tries to quantize a background-independent theory – General Relativity – by means of a necessarily background-dependent method: a conceptual contradiction. It exemplifies that it is not consistently possible to quantize a background-independent theory of spacetime by means of a background-dependent approach, capturing the quantum dynamics of spacetime on (an already fixed) spacetime. – And, to mention it already here, there are at least good reasons to argue that *String Theory* is nothing else but a much more complicated, filigree, byzantine exemplification of this conceptual inadequacy and its implications.[11]

---

[8] However, this strategy for the development of a theory of Quantum Gravity, i.e. constructing it by means of a (direct) quantization of General Relativity, intended to identify and capture the quantum properties of gravity and spacetime, will only be successful if gravity is indeed a fundamental interaction, if the gravitational field (as well as spacetime) has indeed quantum properties. See section 2.2.
[9] Cf. DeWitt (1967a, 1967b).
[10] All other interactions couple only to their 'charges', not to energy.

*Canonical Quantization*

The strategy as well as the problems of the *Canonical Quantization* of General Relativity are completely different from those of the *Covariant Quantization* approach. It is a much more sophisticated, intrinsically non-perturbative, background-independent, full-blown quantization of General Relativity, starting from its Hamiltonian formulation. Nonetheless the old geometrodynamical[12] *Canonical Quantization* approach, which started from a quantization of a Hamiltonian formulation of General Relativity with the metric and the curvature of spacetime as basic variables, led to severe and probably insoluble problems. Its fundamental equation, the *Wheeler-Dewitt equation* (i.e. the quantized counterpart to the *Hamiltonian constraint* of the classical theory, which captures the temporal aspect of its diffeomorphism invariance), turned out to be ill-defined and led to severe conceptual and mathematical problems.

A possibly more successful reincarnation of the *Canonical Quantization* approach is *Loop Quantum Gravity*.[13] It starts from a Hamiltonian formulation of General Relativity based on the *Ashtekar variables*[14] (a spatial SU(2) connection variable and an orthonormal triad) instead of the metric and the curvature of spacetime as basic variables. *Loop Quantum Gravity* uses the Dirac quantization method[15] for constrained Hamiltonian systems. After quantization, one finds (already on the kinematical level of description) a discrete, polymer-like graph structure for the spatial hypersurfaces; this *spin network* structure represents the discrete eigenvalues of two geometric operators one can define in *Loop Quantum Gravity*: the *area* and the *volume operator*.

But, a severe problem remains:[16] No low-energy approximation and no classical limit have been derived as yet. It has not been possible to reproduce the known low-energy phenomenology of gravity or to derive the Einstein field equations (or anything similar to them) as a classical limit. – And here it should be emphasized that it is not a necessary requirement for a theory of Quantum Gravity to quantize General Relativity in a conceptually coherent way (although this might be a natural strategy). Rather, the basic and indispensable requirement for such a theory is that it is able to reproduce the phenomenology of gravity: the classical, low-energy case. Should it not be possible to do that in *Loop Quantum Gravity*, this would be its end.[17] So it seems at least to be reasonable to look for alternatives.

---

[11] See section 3.
[12] Cf. DeWitt (1967), Kuchar (1986, 1993), Ehlers / Friedrich (1994).
[13] Cf. Ashtekar (2007, 2007a), Ashtekar / Lewandowski (2004), Rovelli (1997, 2004), Thiemann (2001, 2002, 2006), Nicolai / Peeters (2006), Nicolai / Peeters / Zamaklar (2005). For a literature survey see Hauser / Corichi (2005).
[14] Cf. Ashtekar (1986, 1987). The basic variables are modified again in *Loop Quantum Gravity*: from Ashtekar's connection variables to loop variables (Wilson loops).
[15] Cf. Henneaux / Teitelboim (1992). The Dirac quantization method consists in a quantization of the full, unconstrained Hamiltonian phase space of the classical theory – canonical commutation relations for the quantum counterparts of the classical variables, an operator algebra, and finally, the quantum counterparts of the classical constraints are to be defined – with the intention to take the quantum constraints into account (to 'solve the constraints') afterwards, and to identify thereby the true physical states.
[16] There are many more problems. Especially, it was not possible, as yet, to solve the quantum Hamiltonian constraint (i.e. the Wheeler-DeWitt equation) completely. Not even the definition of the quantum Hamiltonian constraint is unambiguous.
[17] Should *Loop Quantum Gravity* instead finally be able to master these problems and succeed in the reproduction of the phenomenology of gravity, it is already clear that it has pretty radical implications in comparison to the well-established theories of physics. Its dynamics does not fulfill unitarity and all observables are non-local. The probably most radical

## 2.1.2. Other Strategies for the Identification of Quantum Properties of Gravity

If no theory that can be constructed by means of a (direct) quantization of General Relativity should be able to identify the possible quantum properties of gravity (and spacetime) and at the same time to reproduce General Relativity (or at least its phenomenology) as a classical limit or a low-energy approximation (up to the exactitude of the already existing empirical data), two alternative options remain: One could either try to find a quantum theory with the appropriate classical limit by means of the quantization of another classical theory instead of General Relativity. Or one could try to construct or to find such a theory 'directly': without any quantization of a classical theory at all. – There are no instantiations of the latter option (at least if one looks for theories that try to identify quantum properties of gravity and/or spacetime[18]), but – as we will see – of the former one: *String Theory*.[19]

## 2.2. Gravity as a Residual, Intrinsically Classical Phenomenon – The Search for the Quantum Substrate

Strategies for the development of a theory of Quantum Gravity that try to identify the quantum properties of gravity are only adequate, if gravity has indeed quantum properties. But contrary to this assumption – and the intuitions behind it – the possibility can not be excluded that gravity is an intrinsically classical phenomenon. In this case – at least if Quantum Mechanics should be universally valid – gravity can not be a fundamental interaction; it would have to be an induced or residual interaction, caused by non-gravitational interactions and their corresponding degrees of freedom.[20] It would have to be an emergent phenomenon, resulting from a quantum substrate without gravitational degrees of freedom. Only then, there would be no conflict with the arguments against semi-classical theories, because, on the fundamental level, there would be no semi-classical hybrid dynamics that leads to conceptual inconsistencies; there would only be the quantum substrate, governed by fundamental quantum interactions to which gravity would not belong. A theory describing the dynamics of the gravitational field would then be an effective theory describing the intrinsically classical dynamics of collective degrees of freedom that result from a completely different quantum substrate; this classical theory would have to be recovered from the fundamental theory by means of something like a statistical approximation over the (more) fundamental degrees of freedom of the substrate.

In this case, theories that try to identify quantum properties of gravity would be conceptually and empirically inadequate. And it would be completely nonsensical to quantize gravity. There would be no quantum properties of gravity, no gravitons etc. Gravity would be based on a substrate without any gravitational degrees of freedom. A quantization of gravity would correspond to a quantization of collective, macroscopic degrees of freedom. A quantization of General Relativity would be the quantization of an effective theory describing the dynamics of these collective degrees of freedom. It would be as useful as a quantization of the Navier-Stokes equation of hydrodynamics. The resulting 'Theory of Quantum Gravity' would be analogous to something like a Quantum

---

of its consequences is known as the *problem of time*; cf. Belot / Earman (1999, 2001), Butterfield / Isham (1999), Isham (1993), Kuchar (1991, 1992), Rickles (2006), Rovelli (1991, 2002, 2007), Unruh / Wald (1989).

[18] For approaches that do not subscribe to this goal, see section 2.2.

[19] See section 3.

[20] As a first idea with regard to this emergence of gravity, one could think possibly of an analogy to the emergence of *Van der Waals forces* from electrodynamics.

Hydrodynamics: an artificial, formal quantization of a classical theory capturing collective, macroscopic degrees of freedom, without any implications for, or any clarifications with regard to, an underlying quantum substrate. It would be simply the wrong degrees of freedom which are quantized. – An adequate strategy would instead consist in a search for the substrate dynamics from which gravity emerges. 'Quantum Gravity' would then be the label for a theory that describes this substrate and that explains how gravity emerges from this substrate.[21]

If gravity should indeed be an intrinsically classical, residual or induced, emergent phenomenon, without any quantum properties, what about spacetime? – If we see General Relativity as an adequate effective description of classical gravity, the general relativistic relation between gravity and spacetime, i.e. the geometrization of gravity, should be taken seriously (at least as long as no better reasons make this questionable). General Relativity would have to be seen as a classical, low-energy, long-distance limit to a searched-for theory describing the quantum substrate from which gravity *and* spacetime results. This substrate would neither contain gravity, nor would it presuppose spacetime, at least not the continuous, dynamical spacetime of General Relativity, into which the gravitational field is encoded as metric field. The spacetime of General Relativity – we would have to expect – would be, like gravity, an emergent phenomenon. It would not be fundamental, but the macroscopic result of the dynamics of a non-spacetime ('pregeometric'[22]) substrate.

Under these conditions, the decisive questions of 'Quantum Gravity' are the following: From which structure do gravity and/or spacetime emerge? Of what entities and interactions does the substrate consist? Does matter (and do other quantum fields) also emerge from the substrate? – Meanwhile, there exist a lot of different, more or less (mostly less) convincing scenarios that try to answer these questions; some are conceptually interrelated and some are completely independent. They differ especially with regard to their specific construction of the substrate dynamics from which spacetime and/or gravity is supposed to emerge. There are hydrodynamic and condensed matter approaches[23], pregeometric quantum state space scenarios[24], thermodynamical approaches[25], computational and information theoretical approaches[26], etc. Some of these scenarios take General Relativity as an

---

[21] So, if gravity should be a non-fundamental, emergent, intrinsically classical phenomenon, this changes the motivations for a theory of Quantum Gravity significantly. A conceptual incompatibility resulting from classical gravity in a quantum world does not exist any more in this case. Resulting as a classical phenomenon from a quantum substrate, gravity would already by compatible with the quantum. A theory of 'Quantum Gravity' would not have to make gravity compatible with Quantum Mechanics, but to explain the emergence of gravity from the quantum substrate.

[22] 'Pregeometric' does not necessarily mean 'non-geometric', but 'pre-general-relativistic-spacetime-continuum'.

[23] Cf. Volovik (2000, 2001, 2003, 2006, 2007, 2008), Hu (2005), Hu / Verdaguer (2003, 2004, 2008), Finkelstein (1996), Zhang (2002), Tahim et al. (2007), Padmanabhan (2004), Eling (2008), Sakharov (2000), Visser (2002), Barcelo / Liberati / Visser (2005). – Actually, it is unclear at the moment, to what extent the hydrodynamic and condensed matter models are in conflict with basic conceptual implications of General Relativity, e.g. what kind of background they need, and if they necessarily need an external time parameter or a quasi-local change rate.

[24] Cf. Kaplunovsky / Weinstein (1985); see also Dreyer (2004). The Kaplunovky-Weinstein scenario, based on standard Quantum Mechanics, presupposes an external time parameter, which is at least incompatible with General Relativity. However, first ideas with regard to the question how a temporal dynamics could emerge from a timeless 'dynamics' are arising; cf. Girelli / Liberati / Sindoni (2008).

[25] Cf. Jacobson (1995, 1999), Eling / Guedens / Jacobson (2006), Jacobson / Parentani (2003). See also Padmanabhan (2002, 2004, 2007). Jacobson has shown that the Einstein field equations can be derived from a generalization of the proportionality between entropy and horizon area for black holes (*Bekenstein-Hawking entropy*). For that, one needs the thermodynamical relations between heat, temperature and entropy. Temperature has to be interpreted as Unruh temperature of an accelerated observer within a local Rindler horizon. Heat is to be interpreted as energy flow through a causal horizon in the past, leading to a curvature of spacetime, corresponding to a gravitational field.

[26] Cf. Lloyd (1999, 2005, 2007), Hsu (2007), Livine / Terno (2007), Zizzi (2001, 2004, 2005), Hardy (2007), Cahill (2002, 2005), Cahill / Klinger (1996, 1997, 1998, 2005), Requardt (1996, 1996a, 2000), Wheeler (1979, 1983, 1989). One of the advantages of the idea that spacetime could be an emergent information-theoretical phenomenon is that some

adequate description of gravity and spacetime – as an effective theory for the macroscopic, low-energy regime –, keep to the general relativistic relation between gravity and spacetime, and treat them as emerging together from a pregeometric substrate. Others take General Relativity as a theory with limited validity, even for the classical, macroscopic regime – especially with regard to its geometrization of gravity –, and describe the emergence of gravity from a substrate that already presupposes spacetime. Some are pregeometric with regard to space, but not with regard to time, which is presupposed, either as a continuous parameter, or in form of discrete time steps. Most of the scenarios presuppose the validity of Quantum Mechanics on the substrate level, but a few[27] try also to explain the emergence of Quantum Mechanics from a (in some cases deterministic) pre-quantum substrate.

An approach that goes (sometimes) by the name *Quantum Causal Histories*[28] can possibly be seen as one of the most clear-cut, almost paradigmatic attempts to formulate a pregeometric, quantum-computational theory of Quantum Gravity.[29] In this approach, it is possible to show in a conceptually consistent way how gravity and spacetime could – in principle – emerge as intrinsically classical phenomena from a pregeometric quantum substrate[30] that does not have any gravitational degrees of freedom at all. In this scenario, macroscopic spacetime and classical gravity do not result from a coarse-graining of quantum-geometric degrees of freedom – those do not exist according to the *Quantum Causal Histories* approach –, but from the dynamics of (emergent) propagating coherent excitation states[31], resulting (and at the same time dynamically decoupled) from the substrate

---

of the problematic implications of the hydrodynamic and condensed matter models, e.g. their possible inability to achieve background-independence, can be avoided.

[27] Cf. Cahill (2002, 2005), Cahill / Klinger (1996, 1997, 1998, 2005), Requardt (1996, 1996a, 2000).

[28] Cf. Markopoulou (2000, 2000a, 2000b, 2004, 2006, 2007), Dreyer (2004, 2006, 2007) (Dreyer calls his approach *Internal Gravity*), Kribs / Markopoulou (2005), Konopka / Markopoulou / Smolin (2006) (*Quantum Graphity*), Konopka / Markopoulou / Severini (2008), Hawkins / Markopoulou / Sahlmann (2003).

[29] *Quantum Causal Histories* can not only be seen as the paradigmatic case of a pregeometric theory of Quantum Gravity, but also as a synthesis or a point of convergence of many different approaches to a pregeometric quantum substrate. They are, on the one hand, a conceptual extension of Sorkin's *Causal Set* approach (Cf. Bombelli / Lee / Meyer / Sorkin (1987), Sorkin (2003), Rideout / Sorkin (2000, 2001), Rideout (2002), Henson (2006), Surya (2007)), enriched by the *Holographic Screens* idea (Cf. Markopoulou / Smolin (1999)) and elements from Lloyd's *Computational Universe* scenario (Cf. Lloyd (1999, 2005, 2007)), which itself owes a lot to Wheeler's *It from bit* (Cf. Wheeler (1989)). On the other hand, *Quantum Causal Histories* can also be seen as a generalization of causal spin networks and of the *Spin Foam* approach (Cf. Oriti (2001, 2003), Livine / Oriti (2003), Perez (2003, 2006), Baez (1998, 2000), Markopoulou / Smolin (1997)), enriched by elements from *Algebraic Quantum Field Theory*.

[30] The basic assumptions of the *Quantum Causal Histories* approach with regard to this substrate are the following:
– Causal order is more fundamental than properties of spacetime (like metric or topology).
– Causal relations are to be found on the substrate level in form of elementary causal network structures.
– There are no (continuous) spacetime degrees of freedom on the substrate level.
– Only a finite amount of information can be ascribed to a finite part of the substrate network of causal relations.
– Quantum Mechanics is valid on the fundamental level.
The substrate is modeled as a relational network of quantum systems with only locally defined dynamical transitions. The basic structure is a discrete, directed, locally finite, acyclic graph. To every vertex (i.e. elementary event) of the graph a finite-dimensional Hilbert space (and a matrix algebra of operators working on this Hilbert space) is assigned. So, every vertex is a quantum system. Every (directed) line of the graph stands for a causal relation: a connection between two elementary events; formally it corresponds to a quantum channel, describing the quantum evolution from one Hilbert space to another. So, the graph structure becomes a network of flows of quantum information between elementary quantum events. *Quantum Causal Histories* are information processing quantum systems; they are quantum computers.

[31] According to the *Quantum Causal Histories* approach, it is topology that stabilizes these coherent excitation states. The idea is that they can be identified with stable topological knot structures: braids with crossings and twists. These topological structures seem to be conserved by the substrate dynamics because of topological symmetries, i.e. because of corresponding topological conservation principles. Cf. Bilson-Thompson / Markopoulou / Smolin (2006), Bilson-Thompson (2005).

dynamics.[32] These coherent excitation states give rise to spacetime, because they behave dynamically (and especially with regard to the symmetries and invariances of this dynamics) as if they were living in a spacetime. And, as these coherent excitation states, because of their characteristic intrinsic properties, can be interpreted as matter degrees of freedom,[33] the emergence of spacetime and of gravity is intrinsically coupled to the emergence of matter. The spacetime of the *Quantum Causal Histories* approach is nothing more than an implication of the dynamical behavior of (emergent) matter. Spacetime is here a completely relational construct, an expression of the phenomenology of matter dynamics. And the matter degrees of freedom give at the same time rise to gravity, because the spacetime they bring forth by means of their behavior is a curved spacetime.[34] Gravity is nothing more than an expression of this curved spacetime.[35]

## 3. String Theory

In the context of the different strategies to construct a theory of Quantum Gravity, *String Theory*[36] can be seen as an approach that tries to eliminate conceptual incompatibilities between General Relativity and Quantum Mechanics by means of capturing (assumed) quantum properties of gravity within the (slightly extended) model-theoretical paradigm of (perturbative) Quantum Field Theory.[37] Like the unsuccessful *Covariant Quantization* approach – the other approach to Quantum Gravity that uses essentially the same model-theoretical apparatus – it tries to capture gravity in form of an exchange of gravitons[38]. But, in contrast to this approach, *String Theory* is not only not non-renormalizable; it seems to be finite. Obviously, this is a consequence of the main conceptual differences between the *Covariant Quantization* approach and *String Theory*. On the one hand, *String Theory* does not try to reach at a quantum description of gravity by means of a (direct) quantization of General Relativity; instead, it can be seen as the result of a quantization of a different classical dynamics – that of a relativistic, one-dimensionally extended object: the 'string'. And, on the other hand, what seems to be crucial in avoiding non-renormalizability: *String Theory* does not try to capture the quantum dynamics of gravity in an isolated way, but by means of a nomologically unified description of all fundamental interactions[39] – including gravity (and thereby taking gravity as a fundamental interaction).

---

[32] Cf. Kribs / Markopoulou (2005).

[33] Interestingly, the basic properties of these stable topological structures can be identified with the well-known intrinsic properties of elementary particles. E.g., the twist of a braid structure can be interpreted as electromagnetic charge. There are also topological counterparts to charge conjugation, to quark colors, to parity, etc. (Cf. Bilson-Thompson / Markopoulou / Smolin (2006), p. 6.) In this manner, all particles of the Standard Model can be identified with specific topological structures. (Cf. Bilson-Thompson / Markopoulou / Smolin (2006), Bilson-Thompson (2005), Bilson-Thompson / Hackett / Kauffman / Smolin (2008).) Naturally, this spectrum of topological structures does not contain any counterpart to the graviton. According to the *Quantum Causal Histories* approach there are no gravitons. Gravity is an intrinsically classical, emergent phenomenon; it does not have any quantum properties or quantum constituents. – However, what is still missing, is a dynamical explanation that elucidates the identification of the basic properties of the stable topological structures with the intrinsic properties of elementary particles. It should, finally, be possible to derive energy conservation principles from the dynamics of the stable topological structures (which should be translation-invariant); and this should, not at least, lead to an explanation for particle masses.

[34] There are already concrete indications for a curved spacetime with Lorentz signature. The still unproved central hypothesis of the *Quantum Causal Histories* approach is that the Einstein field equations are necessarily an implication of the dynamics of the coherent excitation states, and that they can finally be derived from the substrate dynamics. Cf. Markopoulou (2007) p. 19.

[35] And gravity has a finite propagation speed because the coherent excitation states of the substrate, the matter degrees of freedom, have a finite propagation speed.

[36] Cf. Polchinski (2000, 2000a), Kaku (1999), Green / Schwarz / Witten (1987). For a survey of the literature, see Marolf (2004).

[37] Initially, *String Theory* was not developed for this purpose. See section 3.2.

[38] Gravitons are, in contrast to the *Covariant Quantization* approach, no basic constituents in *String Theory*. See section 3.2. and 3.3.

## 3.1. Nomological Unity of the Forces

The idea of a substantial unity of nature is very old. It leads back to the *arché* concept of pre-socratic philosophy of nature. A (theoretically and mathematically reinforced) reincarnation of this metaphysical idea forms, together with ontological reductionism, the core of one of the most ambitious strategic programs to be found in modern physics: the program of a *nomological unification* of all fundamental interactions.[40] – A more moderate variant tries to realize at least a *conceptual unification*, to be achieved by the (complete) elimination of (conceptual) contradictions between (fundamental) physical theories.

There are many successful examples of conceptual as well as of nomological unifications in the history of physics. With Newtonian physics the old separation into terrestrial and celestial mechanics was overcome. With Maxwell's electrodynamics the nomological unification of electricity, magnetism and optics could be achieved. The Einsteinian theories of relativity established the conceptual compatibility of mechanics, classical electrodynamics and, finally, gravitation. With the Standard Model of elementary particle physics we have a theoretical construct at our disposal which achieves, at least conceptually[41], the unification of the electromagnetic, weak, and strong forces. – So, apparently, only the problem of Quantum Gravity remains. And one of the central assumptions of *String Theory* is that this problem can (only) be solved by means of a *nomological unification* of all fundamental forces (taking gravity to be one of these).

String theorists often refer to the *Glashow-Salam-Weinberg* (GSW) *model* of electroweak interactions as a motivation for this assumption. And indeed, the GSW model is one of the best examples of a successful nomological unification. However, one has to remember that the success of the GSW model did not simply consist in its realization of a nomological unification of electromagnetic and weak interactions or in the fact that it was the first renormalizable theory to describe weak interactions; rather, its success became manifest in the fact that it made empirically testable numerical predictions which, soon afterwards, were confirmed by experiments at CERN. No one, not even its inventors, took the GSW model seriously only because it was a unified theory, but because of its empirical success. – And this is the crucial difference to *String Theory* and its nomological unification strategy: In contrast to the GSW model, *String Theory* does not make any quantitative, numerical predictions, necessary for an independent empirical test. As long as this problem exists, there is not even the possibility of a direct empirical corroboration at all. Therefore, *String Theory* can hardly be compared to (or even seen on the same level with) empirically successful nomological unifications like the GSW model. Without any direct empirical corroboration, the idea of a complete unity of all (fundamental) forces remains hypothetical. It remains a mere metaphysical idea as long as we don't have an extensive, really convincing empirical corroboration of a theory which achieves a nomologically unified and at the same time conceptually consistent description of all interactions. And this presupposes empirically testable quantitative predictions of such a theory.

---

[39] It seems to be the existence of all the different string oscillation states that makes the perturbative expansions not only renormalizable, but even finite. Different divergent contributions to the expansion seem to cancel out each other. But this does only work if one assumes supersymmetry (i.e. the symmetry between bosonic and fermionic states) for the string dynamics. See sections 3.2. and 3.3.
[40] Cf. Weinberg (1992).
[41] Attempts to develop a nomologically unified description of these three interaction, a *Grand Unified Theory*, were unsuccessful. See below in this section.

So, that a theory is nomologically unified is per se no argument supporting the theory, as long as nomological unity is nothing more than a metaphysical idea. – However, in the context of Quantum Field Theory, there exists at least one specific physical argument that supports the idea of a nomological unity of the forces. It comes from the (extrapolated) convergence of the coupling constants of the electromagnetic, the weak, and the strong interactions with increasing energies. But it is highly questionable that gravity can really be included in the quantum field theoretical picture indicating this convergence.[42] So, it remains also highly questionable if the convergence argument is of any use for *String Theory* and its assumption that a nomological unification leads to a consistent theory of Quantum Gravity.

And here, one should remember the negative counterpart to the sequence of successes of the nomological unification program: the history of its failures. Einstein's futile attempt at a formulation of a *Unified Field Theory* is certainly one of the most prominent examples. But one should also remember the *Grand Unified Theories*, developed to achieve a nomological unification of electroweak and strong interactions; this step beyond the (only conceptually, but not nomologically unified) Standard Model led, not at least, to contradictions between numerical predictions for proton decay and the relevant empirical data; and it ended with a great disillusionment in high energy physics.

So, one can ask, what are the reasons for these failures of the nomological unification program? Could it be that the idea of a thoroughgoing unity of nature is simply wrong? Or does not even a complete unity of nature guarantee the success of the nomological unification program in physics? Does not even a complete unity of nature guarantee an all-encompassing and adequate description of nature that reflects this unity? There might be limits to our epistemic ambitions, making this impossible.[43] Although there exists a long sequence of successes in the past, from Newton to the GSW model, the nomological unification program could well turn out to be finally inadequate and even disadvantageous for the future prospects of physics. Metaphysical motivations alone do not necessarily lead to any descriptive success in the empirical sciences. – Quite independently from these more fundamental skeptical doubts, a nomologically unified theory has, at least, to be conceptually consistent (like any physical theory); and it has to be physically motivated with regard to its central conceptual ideas.

## 3.2. How It All Began

As its historical development easily elucidates, it is *not* because of an initial decision that such a strategy would lead to an adequate theory of Quantum Gravity that *String Theory* turned out to be a nomologically unified approach. Originally, in the late sixties, *String Theory* started from a devel-opment in the context of hadron physics. It was the casual discovery of a correspondence between hadron scattering behavior and Euler's beta-function which led Veneziano in 1968 to a model which could be identified in 1970 by Susskind, Nielsen, Nambu and Goto as describing the quantized dynamics of a relativistic string. But this 26-dimensional bosonic string theory was not very successful in the context of hadron physics and lost its relevance after the rise of Quantum Chromodynamics and the quark model. – Then, surprisingly, *String Theory* experienced a sudden reincarnation as a candidate theory for Quantum Gravity. This reincarnation was essentially triggered by the discovery, made by Scherk and Schwarz in 1974, that the quantization of the dynamics of a relativistic string leads, besides many other states, to spin-2 states, completely useless in hadron physics. It was this discovery and the interpretation of these spin-2 states – after a shift in the relevant energy scale of about 15 powers of ten[44] – as gravitons which finally determined the actual role of *String Theory*. After this shift of the energy level of the oscillating string (the string tension) to the Planck level, and after the elimination of various intratheoretical anomalies, it was possible to reproduce, at least formally, the Einstein field equations as a classical limit.[45] *String Theory* mutated from an unsuccessful theory in hadron physics to a prospective unified theory of all interactions, including gravity. This transmutation could be made complete not at least because of an additional shift to a supersymmetric formulation, now – forced again by anomaly elimination procedures – with a ten-dimensional target space (instead of the 26 dimensions of the old bosonic theory), which made it possible to include also fermionic matter fields. – From here all further developments started: the discovery of five perturbative, supersymmetric, ten-dimensional string theories; attempts to construct four-dimensional low-energy implications of these theories by means of e.g. compactification and orbifold techniques, aimed to be compatible (or at least comparable) with low-energy phenomenology, but not even leading to any numerical predictions; ideas about possible mechanisms for supersymmetry breaking; the discovery of duality relations between the different perturbative string theories (and the eleven-dimensional *Supergravity* theory, leading to ideas about an eleven-dimensional non-perturbative, still non-existing extension of *String Theory*, called *M-Theory*); the discovery of branes and other objects in the indirectly extrapolated non-perturbative regime of the thereby slightly extended perturbative string theories; still unsuccessful attempts to develop a fully non-perturbative formulation of *String Theory*; the discovery that *String Theory* obviously does not lead to a unique description, but to a plethora of nomologically, physically and phenomenologically different low-energy scenarios – scenarios with different symmetries, parameter values, values of the cosmological constant, etc. – known as the *String Landscape*,[46] where *String Theory* finally loses any prospect of having any (maybe still hidden) predictive power and becomes open to ideas of an anthropic selection; etc.

---

[42] See sections 2.1.1. and 3.3.

[43] Reality could well turn out to be a patchwork of rather independent phenomenological areas, at least when we try to describe it by means of our epistemic capacities. It could be something which can not be described with empirical adequacy by means of a coherent, unified, fundamental physical theory, but only by a collection of effective theories which would find their relevance in a direct and close coupling to specific phenomenal areas. In this case, even if we would be able to unify by force, the results of this unification would not describe reality; rather, it would be a mere conceptual extrapolation without descriptive content. – Consequently, in philosophy of science, there can be found a widespread scepticism (i) with regard to the philosophical motivations behind the idea of a unity of nature as well as (ii) with regard to its fertility for science and its descriptive aims. Cartwright is probably the most prominent of those who doubt that nature can be adequately described by means of fundamental and nomologically unified physical theories. (Cf. Cartwright (1994, 1999; also 1983, 1989).) According to her view, reality is something which can only be captured approximately, by a patchwork of effective theories which have only a limited reliability for a specific context. These effective theories will not even have to be completely compatible to each other. So, even the requirement of a conceptual, model-theoretical uniformity and consistency might be to strong.

*String Theory* is obviously not the result of a planned strategy to develop a theory of Quantum Gravity or a nomologically unified description of all interactions; rather, it fell into this role by means of a casual discovery. It is not the result of a planned development of a theory aimed at the description of a certain context of relevance or at the solution of specific physical problems, but rather that of a casual finding of a theory: a mathematical construct which, after having lost its originally intended physical relevance, found by chance a new context of relevance, even without looking for it. Then, in its new role as a realization of an all-encompassing nomological unification of the forces, all the motivations for the conceptual as well as for the nomological unification program could be claimed. These were, on the one hand, the problems resulting from the mutual con-ceptual incompatibility between General Relativity and Quantum Mechanics, forming the motivational core of the search for a theory of Quantum Gravity. On the other hand, the still vivid memory of the former success of the nomological unification of electromagnetic and weak interactions (the *Glashow-Salam-Weinberg model*) lead now to the idea of having found – here in the more extended context of a nomologically unified theory of all interactions – a solution for the virulent problems with a quantization of gravity. And *String Theory* had no other option: Its only possible justification of existence as a physical theory lies in the unification program. The existence of gravitons as well as spin-1 gauge bosons makes only sense, if *String Theory* plays the role of an all-encompassing, nomologically unified theory of all interactions. – But, is *String Theory* indeed able to play this role successfully?

---

[44] *String Theory* reproduces general relativity (as well as gauge invariances, possibly those of the standard model of quantum field theory) as a low-energy approximation. But, general relativity comes only with the phenomenologically correct parameter values, if the string length and tension are assumed to lie in the order of magnitude of the Planck scale.

[45] That it is possible to reproduce the Einstein equations from *String Theory* does not necessarily mean that it reproduces General Relativity in a full-blown sense. See section 3.3.2.

[46] Cf. Banks / Dine / Gorbatov (2004), Dine (2004), Douglas (2003), Susskind (2004, 2005). See also Hedrich (2006).

## 3.3. The Epicycles of Quantum Field Theory

It is not only with regard to its role as a nomologically unified approach to Quantum Gravity that *String Theory* had no other option. There a some further 'decisions' without alternative at the core of the theory. So, *String Theory* did not make the choice for a higher-dimensional target space because we are living obviously within a ten- or whatever-higher-dimensional spacetime. And, *String Theory* did not make the decision to take supersymmetry into its structural repertoire because nature is obviously supersymmetric. These 'decisions' were not made out of physically motivated reasons. Rather, they are the result of a virtually self-developing mathematical construct, driven almost exclusively by internal consistency requirements.[47] The most essential component of this virtually self-developing mathematical construct consists in the (only slightly extended) model-theoretical apparatus of perturbative Quantum Field Theory. And *String Theory* did not deliberately chose this model-theoretical apparatus out of a spectrum of possible model-theoretical bases. It simply used it from the beginning until now, never discussing its conceptual adequacy. But here, in the context of Quantum Gravity, were the application of this model-theoretical apparatus leads to ever more bizarre implications, the possibility should be taken into account that this apparatus might simply meet its limits of physical applicability.[48]

### 3.3.1. Consistency Requirements, Theoretical Artifacts and the Dominance of Internal Problems

That *String Theory* is foremost a self-developing mathematical construct, driven in its development almost exclusively by consistency requirements originating within this construct, finds its direct expression in the dominance of *internal* (intratheoretical, conceptual, model-theoretical and mathematical) over *external* (real physical) problems.[49] One could say that *String Theory*, at least since its change of context (and strategy) from hadron physics to Quantum Gravity, seems to be by far too much preoccupied with its internal conceptual and mathematical problems to be able to find concrete solutions to relevant external physical problems. Internal problems led with their attempted solutions to further internal problems, and so on. The result of the successively increasing self-referentiality is a more and more enhanced decoupling from phenomenological boundary conditions and necessities. The contact with the empirical does not increase, but gets weaker and weaker. The result of this process is a labyrinthine mathematical structure with a completely unclear physical relevance. Often it is not even obvious if certain components of *String Theory* are to be seen as part of the same picture, or as part of complementary formulations, or as part of competitive alternatives or scenarios.

---

[47] To make that clear: Consistency requirements, even if some basic phenomenological constraints are taken into account, are not sufficient to establish an adequate theory of Quantum Gravity – especially if there are alternatives.
[48] Apart from the bizarreness of *String Theory*, there are independent arguments against the conceptual adequacy of this model-theoretical apparatus in Quantum Gravity. See section 3.3.3.
[49] For a more extended discussion of *String Theory* and its problems, see Hedrich (2002, 2006, 2007, 2007a).

One of the most basic of the internal problems results from the fact that there are (at least) five distinct perturbative string theories – and not only one. By means of relations (dualities) between these five perturbative theories, string theorists try to establish a non-perturbative framework (that, at the same time, should possibly solve the problem of the background-dependence[50] of the perturbative string theories). But, as yet, no consistent non-perturbative, analytical framework exists (not even a background-dependent one).

A further internal problem follows from the fact that (physically relevant) perturbative string theories have necessarily to be supersymmetric. It is *not* the fact that empirical indications for supersymmetry were found, that force consistent string theories to include supersymmetry. The arguments of supersymmetry are much more indirect: Without supersymmetry, *String Theory* has no fermions and no chirality; instead there are tachyons that make the vacuum unstable. And supersymmetry has certain additional conceptual advantages: it leads very probably to the finiteness of the perturbation series, thereby avoiding the problem of non-renormalizability which haunted former perturbative attempts at a quantization of gravity; and there is a close relation between supersymmetry and Poincaré invariance which seems reasonable for Quantum Gravity. But, nonetheless, there are no empirical indications for supersymmetry, and not all conceptual advantages are necessarily part of nature – as the example of the elegant, but empirically unsuccessful *Grand Unified Theories* demonstrates. We do not see supersymmetry in our world, at least not an unbroken one. So, *String Theory* should be able to explain why we do not see supersymmetry, although it is a necessary ingredient of the theory. And there should be numerical predictions with regard to an obviously broken supersymmetry. But *String Theory* has a lot of problems with a broken supersymmetry; and it does not lead to any quantitative predictions at all. One of the reasons for that is that the problems with regard to supersymmetry breaking seem to be coupled to another internal problem: that of explaining the transition from the necessarily ten-dimensional dynamics of the string (forced again by internal consistency requirements) to the four-dimensional phenomenology of our world. And again: It is *not* the fact that we are obviously living in a ten-dimensional world which forces *String Theory* to a ten-dimensional description. It is that (supersymmetric) perturbative string theories are anomaly-free only in ten dimensions; and they contain gravitons only in a ten-dimensional formulation. The resulting question, how the four-dimensional spacetime of phenomenology comes off from ten-dimensional perturbative string theories (or its possibly eleven-dimensional non-perturbative extension: the mysterious *M theory* that still does not exist at all, in spite of already having a name), led to the compactification idea and to the braneworld scenarios – and from there to further internal problems. Different proposals for dimensional reduction and compactification mechanisms exist. But even if one takes only one compactification scheme into account, this transition is highly ambiguous; it leads to a plethora of four-dimensional low-energy scenarios with different symmetries, oscillation spectra (boson and fermion spectra), etc.: the *String Landscape*. Although the *String Landscape* seems to consist of $10^{500}$ or more four-dimensional scenarios (theories? / models?), it was not possible to identify at least one resembling or reproducing the low-energy phenomenology of our world, or the dynamical structure (and the symmetries) of the Standard Model respectively. And, there are simply no numerical predictions at all with regard to the masses of the bosonic and fermionic states of the string, not even for one of the many, many string scenarios. – So, shouldn't the *String Landscape* be understood as a clear indication, not only of fundamental problems with the reproduction of the gauge invariances of the Standard Model of elementary particle physics (and the corresponding phenomenology), but of much more severe con-ceptual problems? Almost all attempts at a solution of the internal and external problems of *String Theory* seem to end in the ambiguity and contingency of the multitude of scenarios of the *String Landscape*.[51]

---

[50] See section 3.3.2.

*String Theory* is, all in all, a rather careful adaptation of the mathematical and model-theoretical apparatus of perturbative Quantum Field Theory to the quantized, one-dimensionally extended, oscillating string – with only minimal extensions of this model-theoretical apparatus into the non-perturbative regime for which the declarations of intent exceed by far the conceptual successes. And it is this model-theoretical apparatus that leads to consistency requirements that force the string target space to be higher-dimensional and the string dynamics to be supersymmetrical; these features are, above all, internally motivated, by means of mathematical consistency requirements resulting from the decision to quantize the dynamics of a relativistic string applying the model-theoretical apparatus of Quantum Field Theory; they do not have any independent, truly physical, external motivation. And they are at the same time the origin of most of the further internal problems: problems which came only into existence with the theory under development and its implicit decision for a specific model-theoretical apparatus. They are problems that did not exist before the theory came into existence; they are no original physical problems.

Real physical problems, on the other hand, are of almost no relevance in *String Theory*. Especially, there are no external physical problems that triggered the development of *String Theory*; it found its role as a unified theory of all interactions including gravity by chance, not because it was planned to do so. And there is, at best, only one decisive external physical problem that gives at least a post-hoc motivation for *String Theory*: it is the fact that *String Theory* led, by chance, to the discovery of the graviton, promising thereby a solution to the external physical problem of finding a theory which consistently unifies General Relativity and Quantum Mechanics. But here, *String Theory* does not simply lead to a conceptual unification, an elimination of the conceptual incompatibility between these two theories, i.e. a solution of the original problem at the motivational core of the Quantum Gravity research program; rather it has no other option than to become the realization of a less compelling desideratum within physics: the nomological unification of the forces, achieved here in a conceptually problematic way.[52]

With regard to all other possibly relevant external physical problems and questions for which a theory of Quantum Gravity should try to find a solution or give an answer – there is none. *String Theory* does not lead to any deeper insights with regard to the nature of space and time. Rather it is not even compatible with our already achieved most fundamental insights into the nature of space and time.[53] It does not really lead to an understanding of the possible quantum features of gravitation – if there are any. It does not explain the equivalence of gravitational and inertial mass, or give new answers to the question what *mass* is. – And it scarcely seems to be a far-fetched idea to expect solutions to these problems from a theory of Quantum Gravity. *String Theory* – which promises to give an all-encompassing, nomologically unified description of all interactions – did not even lead to any unambiguous solutions to the multitude of explanative desiderata of the Standard Model of elementary particle physics: the questions with regard to the determination of its specific gauge invariances, broken symmetries and particle generations as well as its 20 or more free parameters, the chirality of matter particles, the reason (and a dynamical explanation respectively) for their masses, etc. Attempts at a concrete solution of these relevant external problems (and explanative desiderata) either did not take place, or they did not show any results, or they led to escalating ambiguities and finally got drowned completely in the *String Landscape*.

---

[51] Cf. Hedrich (2006).
[52] See section 3.2.2.
[53] See section 3.2.2.

The possibly most astonishing example of the irrelevance of fundamental, external (i.e. originally physical) questions and problems and of the almost complete nonexistence of solutions to these problems in *String Theory* is the following: Even after more than three decades of development, there does not exist the slightest idea with regard to a fundamental *physical* principle on which *String Theory* should be based or from which it could be motivated (or even developed). – So, the best one could say is that *String Theory* consists of a sometimes astonishing and sometimes mathematically highly interesting labyrinth of model-theoretical constructs in search for a physically motivated principle which could possibly justify the approach and give it a fundament. Until this happens, the only physical motivation for *String Theory* consists still in the post-hoc discovery that a quantization of a relativistic string leads to spin-2 oscillation states that can be identified with gravitons. That *String Theory* reproduces formally the Einstein equations of General Relativity and the corresponding phenomenology of gravitation is, after this discovery, not much of a surprise; it is a direct implication of the existence of spin-2 states.[54] That the theory, however, reproduces the Einstein equations only formally should become explicit when one takes a look at a further fundamental conceptual problem of *String Theory*.

### 3.3.2. General Relativity, Diffeomorphism Invariance and Background-Independence

One of the most fundamental insights of General Relativity – our empirically well-confirmed classical theory of gravitation and of spacetime – is that it is the metric of spacetime which represents the gravitational field. If we take this geometrization of gravity seriously, that means that the gravitational field is (unlike all other interaction fields) not a field that can be defined on spacetime, but rather a manifestation of spacetime itself; this is reflected formally in the diffeomorphism invariance of General Relativity.[55] Consequently, it is not possible to consistently describe the dynamics of the gravitational field on an already predefined (or even fixed) background spacetime. As long as there are no better, well-founded reasons, a theory of Quantum Gravity has to take into account this background-independence; it has to describe the dynamics of the gravitational field without recourse to an already existing spacetime (metric).[56]

But *String Theory* does not reflect these fundamental insights of the classical theory it is claimed to reproduce. All known formulations of *String Theory* are background-dependent. Like the unsuc-cessful *Covariant Quantization* approach, *String Theory* tries to reconstruct gravity by means of describing an exchange of gravitons on an already presupposed spacetime. The only difference is that gravitons are not fundamental in *String Theory*; they are oscillation states of the string. So, instead of describing the dynamics of (elementary) gravitons on a fixed spacetime, *String Theory* describes the dynamics of one-dimensionally extended strings (with graviton oscillation states) on a fixed spacetime. – And those graviton states (and their post-hoc discovery) are the only reason at all to take *String Theory* into account as an approach to Quantum Gravity.

---

[54] At the time of the discovery of graviton states it possibly wasn't completely clear for string theorists that the dynamics of spin-2 states leads necessarily to the formal reproduction of General Relativity:
   "[...] with appropriate caveats, general relativity is necessarily recovered as the low-energy-limit of any *interacting theory of massless spin-2 particles propagating on a Minkowski background, in which the energy and momentum are conserved [...]."* (Butterfield / Isham (2001) 59)

[55] Because of the diffeomorphism invariance of General Relativity, that can be understood as a gauge invariance (cf. Earman (1986, 1989, 2002, 2006, 2006a), Earman / Norton (1987), Norton (1988, 1993, 2004)), it is physically extremely unreasonable to interpret the spacetime manifold as a substantial entity; the prize for that would consist in rather unmotivated metaphysical assumptions: (i) the negation of *Leibniz equivalence* (i.e. the negation of the identity of the indistinguishable: empirically completely indistinguishable models of spacetime would have to be seen as representations of different spacetimes), and (ii) a completely unmotivated (and unobservable) indeterminism of the theory (as a consequence of the hole argument; cf. the references above). What remains without a substantially interpretable spacetime manifold is: a metric field (identical with the gravitational field; carrying energy and momentum like all other fields), the other interaction fields, the matter fields, and the relations between these fields.

[56] Under extrapolation of the conceptual implications of General Relativity, one could suspect, at least for the time being, that a successful theory of Quantum Gravity will probably not only be a theory describing a *dynamical* spacetime, rather it will, because of the arguments resulting from the diffeomorphism invariance of General Relativity mentioned above, be based on a *relational* conception of spacetime – or it will even lead to an emergent spacetime scenario.

It is easy to see that the idea of a graviton dynamics taking place on an already predefined (fixed) background spacetime is conceptually completely incompatible with the fundamental implications of General Relativity. If gravitons are taken to represent (the quantum dynamics of) the gravitational field, which corresponds to the metric field, then they can scarcely be understood as moving within a (miraculously already existing classical) spacetime. But such an already existing (and even fixed, non-dynamical) spacetime is required by any approach that uses the model-theoretical apparatus of Quantum Field Theory. This apparatus with its fixed background spacetime is fundamentally incompatible with General Relativity and its background-independence (diffeomorphism invariance). It is simply not applicable with regard to the gravitational field, if this field is to be identified with properties of a dynamical spacetime. Here this model-theoretical apparatus reaches its limits of applicability. Taking into account the fundamental implications of General Relativity, gravity can scarcely be seen as an interaction represented simply by field bosons exchanged between matter particles; this is a much too naive picture.

So, one could say that the only motivation at all to take *String Theory* into account as a possible approach to a theory of Quantum Gravity, i.e. the fact that the string has graviton oscillation states, loses ground. At least, the discovery of graviton states is certainly not enough to take the theory seriously as an adequate description of the quantum substrate of gravity and spacetime. *String Theory* tries to reproduce a background-independent theory, General Relativity, and its dynamical implications by means of a fundamentally background-dependent formalism – sticking to the (only slightly extended) model-theoretical apparatus of Quantum Field Theory. Without any further reasonable motivation, such a background-dependent formalism is conceptually inadequate for a theory that claims to describe the quantum properties of gravity and spacetime, and to reproduce General Relativity as a classical limit. Without any excellent idea how a background-independent classical limit could result from a background-dependent quantum theory, the relation between the two theories remains on an exclusively formal level. Without any better reason, a quantum theory leading to General Relativity as a classical limit should be a background-independent theory. The problem was already acknowledged by string theorists; they try to develop a background-independent, non-perturbative formulation of *String Theory* – without success to this day.

### 3.3.3. The Thermodynamics of Black Holes and the Myth of the Continuum

Independently from the background dependence issue, there are further indications of possible conceptual problems with regard to the model-theoretical apparatus of Quantum Field Theory used in *String Theory*. They are coming from the thermodynamics of black holes, a field of considerations that combines implications of General Relativity, Quantum Mechanics, Thermodynamics and Information Theory. The central point of relevance here is that the *Bekenstein-Hawking entropy*[57] of black holes – together with the *Covariant* (or *Holographic*) *Entropy Bound*[58] which transcends in its validity the closer context of the thermodynamics of black holes – leads to clear indications of a finite information content of any spacetime volume: a finite number of physically relevant fundamental degrees of freedom within a bounded spacetime region. This can be taken as an indication *either* of a discrete spacetime structure – if spacetime should be fundamental – *or* of a discrete structure from which spacetime emerges.[59] In any case, it is pointing directly to a substrate with a finite number of true physical degrees of freedom per finite spacetime region – irrespective of the question if spacetime is part fundamental or emergent, i.e. if the fundamental degrees of freedom are (quantum) geometric or 'pregeometric'.[60] So, the implications of the *Bekenstein-Hawking entropy* and of the *Covariant Entropy Bound* are in direct contradiction to the assumption of a continuous spacetime and to the idea of fields defined on such a continuous spacetime; and they are also, especially, in contradiction to the concept of fields that imply an infinite number of degrees of freedom for any spacetime point.

---

[57] Cf. Bekenstein (1973, 1974, 1981, 2000, 2001), Wald (1994, 2001), Bousso (2002).
[58] Cf. Bekenstein (1981, 2000, 2001), Bousso (2002), Pesci (2007, 2008).

Should the substrate indeed be discrete, any continuum description would contain artifacts transcending the true physical degrees of freedom.[61] In physics, the mathematical continuum would then, finally, be nothing more than a mathematical myth. All field-theoretical and quantum field-theoretical approaches, ascribing states (all of which imply themselves the mathematical continuum for their description) to every point of the spacetime continuum, would contain such artifacts that, finally, in the attempt to describe the physical substrate itself – the goal of most of the approaches to a theory of Quantum Gravity – would become completely inadequate. Under these conditions, a failure of approaches that use the model-theoretical apparatus of Quantum Field Theory in the context of an intended substrate description would not be much of a surprise; and, to argue in the reverse direction, such a failure could be seen as an, at least indirect, indication of exactly this fundamental discreteness.

## 4.   Conclusion

Besides the idea of a nomological unity of nature, a metaphysical concept with virulently unclear physical relevance, the most often cited motivation for *String Theory* consists in the fact that the string has a spin-2 oscillation state. *String Theory*, consequently, tries to capture gravity by means of an exchange of gravitons, to be identified with this oscillation state. According to the model-theoretical apparatus of perturbative Quantum Field Theory, a necessary conceptual precondition of the mere idea of graviton exchange, the graviton dynamics takes place on a pre-existing, fixed spacetime. Unfortunately, this is conceptually incompatible with the idea that the gravitational field, captured here by means of this graviton dynamics, is to be identified, according to General Relativity – the theory that *String Theory* allegedly reproduces –, with the metric of a dynamical spacetime. *String Theory* is a background-dependent approach that tries to reproduce a background-independent theory, General Relativity, in a conceptually inadequate way. Gravity – and, a fortiori, the dynamics of spacetime – can not be described in a conceptually consistent way by means of a graviton exchange on a already existing spacetime. The mere concept of a graviton dynamics is conceptually contradictory. So the existence of gravitons as string states can scarcely be a good motivation for taking *String Theory* seriously as an approach to Quantum Gravity.

---

[59] The first alternative – that spacetime has a discrete quantum substructure, i.e. that spacetime has quantum properties leading to a finite information content – finds one of its most direct realizations in the spin networks at the kinematical level of *Loop Quantum Gravity*. But, the, at best, only very limited success of all attempts to quantize gravity and spacetime makes this first alternative less probable. So, the best explanation for the finite information content can be seen in the second alternative; it would then to be read as an indication for a (with regard to its degrees of freedom) finite pregeometric microstructure from which spacetime emerges.

[60] Additional, but much more indirect indications of such a discrete structure come from the singularities that General Relativity predicts (but which transcend its model-theoretical apparatus: differential geometry) and from the divergences that occur in Quantum Field Theory for small distances / high energies. Both could be artifacts of the continuum assumption with regard to spacetime or of the assumption of an infinity of the relevant degrees of freedom respectively. – Interestingly, almost all existing approaches to Quantum Gravity, even those that do presuppose a continuum in their formalism, lead to indications either of a discrete spacetime structure – for those that take spacetime to be a fundamental entity whose quantum properties have to be revealed in the context of a theory that goes beyond General Relativity in exactly this point – or of a discrete ('pregeometric') substrate structure from which spacetime results. What is of specific significance, is that these indications of a discrete substructure are not only present in the more radical approaches, but also in those, like *Loop Quantum Gravity*; that take the fundamentals of General Relativity as well as those of Quantum Mechanics to be essential for a theory of Quantum Gravity. This is most astonishing, because the assumption of a spacetime continuum and of an infinite number of physically relevant degrees of freedom is an inevitable ingredient of General Relativity (differential geometry presupposes the continuum) as well as of Quantum Mechanics and Quantum Field Theory (fields are defined on a classical continuous background space).

[61] *String Theory* would then, at best, be something like a conceptually questionable parameterization of spatially and structurally discrete phenomena carried out by force within the context of the mathematics of the continuum – the inheritance of Quantum Field Theory, if not of the complete traditional model-building in physics.

With regard to the nomological unification program – which is not a constitutive part of the Quantum Gravity program – the situation is not much better: *String Theory* did not even succeed in reproducing the Standard Model of elementary particle physics – in spite of the more than thirty years of its existence, the sequence of metamorphosis it ran through (by means of which *String Theory* has repeatedly cut the ground from under the feet of its critics), and the ever more increasing number of involved physicists. Instead, it lost every prospect of predictive power while sinking, step by step, into a multitude of internal problems. These are resulting in almost all cases from the mathematical and model-theoretical basis chosen at the beginning, leading from higher dimensionality, the inclusion of supersymmetry, the compactification and braneworld scenarios, directly into the abyss of the *String Landscape* where, obviously, every hope with regard to a still hidden and still inactive predictive power of the theory has to be abandoned.

Without being based on any physically motivated fundamental principle, *String Theory* is, at least at the moment, no physical theory at all, but rather a labyrinthine structure of mathematical procedures and intuitions. And, with regard to the goal to describe the searched-for quantum dynamics of the gravitational field, it is very probably the wrong, conceptually inadequate mathematics. Here in the context of Quantum Gravity, the model-theoretical apparatus of Quantum Field Theory reaches its limits of applicability; the attempt to force gravity into the limited spectrum of its conceptual possibilities, leads to further and further epicycles. To explain the existence of gravity in a quantum world, physics is obviously in need of new conceptual ideas and new model-theoretical instruments. Some of these are probably already taking shape.

# 5.  References

Ashtekar, A., 1986: New Variables for Classical and Quantum Gravity, *Physical Review Letters* **57**, 2244-2247
Ashtekar, A., 1987: New Hamiltonian Formulation for General Relativity, *Physical Review* **D 36**, 1587-1602
Ashtekar, A., 2007: An Introduction to Loop Quantum Gravity through Cosmology, arXiv: gr-qc/0702030
Ashtekar, A., 2007a: Loop Quantum Gravity: Four Recent Advances and a Dozen Frequently Asked Questions, arXiv: 0705.2222 [gr-qc]
Ashtekar, A. / Lewandowski, J., 2004: Background Independent Quantum Gravity – A Status Report, *Classical and Quantum Gravity* **21**, R53; also: arXiv: gr-qc/0404018
Baez, J., 1998: Spin Foam Models, *Classical and Quantum Gravity* **15**, 1827-1858; also: arXiv: gr-qc/9709052
Baez, J., 2000: An Introduction to Spin Foam Models of Quantum Gravity and BF Theory, *Lecture Notes in Physics* **543**, 25-94; also: arXiv: gr-qc/9905087
Banks, T. / Dine, M. / Gorbatov, E., 2004: Is there a String Theory Landscape?, *Journal of High Energy Physics* **0408**, 058; also: arXiv: hep-th/0309170
Barcelo, C. / Liberati, S. / Visser, M., 2005: Analogue Gravity, *Living Reviews in Relativity* (Electronic Journal) **8/12**, www.livingreviews.org; also: arXiv: gr-qc/0505065

Bekenstein, J.D., 1973: Black Holes and Entropy, *Physical Review* **D 7**, 2333-2346
Bekenstein, J.D., 1974: Generalized Second Law of Thermodynamics in Black Hole Physics, *Physical Review* **D 9**, 3292-3300
Bekenstein, J.D., 1981: Universal Upper Bound on the Entropy-to-Energy Ratio for Bounded Systems, *Physical Review* **D 23**, 287-298
Bekenstein, J.D., 2000: Holographic Bound from Second Law, arXiv: gr-qc/0007062
Bekenstein, J.D., 2001: The Limits of Information, *Studies in History and Philosophy of Modern Physics* **32**, 511-524; also: arXiv: gr-qc/0009019
Belot, G. / Earman, J., 1999: From Metaphysics to Physics, in: J. Butterfield / C. Pagonis (Eds.): *From Physics to Philosophy*, Cambridge
Belot, G. / Earman, J., 2001: Pre-socratic Quantum Gravity, in: Callender / Huggett (2001)
Bilson-Thompson, S.O., 2005: A Topological Model of Composite Preons, arXiv: hep-ph/0503213
Bilson-Thompson, S. / Hackett, J. / Kauffman, L. / Smolin, L., 2008: Particle Identifications from Symmetries of Braided Ribbon Network Invariants, arXiv: 0804.0037 [hep-th]
Bilson-Thompson, S.O. / Markopoulou, F. / Smolin, L., 2006: Quantum Gravity and the Standard Model, arXiv: hep-th/0603022
Bombelli, L. / Lee, J. / Meyer, D. / Sorkin, R.D., 1987: Space-Time as a Causal Set, *Physical Review Letters* **59**, 521-524
Bousso, R., 2002: The Holographic Principle, *Reviews of Modern Physics* **74**, 825-874; also: arXiv: hep-th/0203101
Butterfield, J. / Isham, C., 1999: On the Emergence of Time in Quantum Gravity, in: J. Butterfield (Ed.): *The Arguments of Time*, Oxford, 111-168; also: arXiv: gr-qc/9901024
Butterfield, J. / Isham, C., 2001: Spacetime and the Philosophical Challenge of Quantum Gravity, in: Callender / Huggett (2001); also: arXiv: gr-qc/9903072
Cahill, R.T., 2002: Process Physics – From Quantum Foam to General Relativity, arXiv: gr-qc/0203015
Cahill, R.T., 2005: *Process Physics – From Information Theory to Quantum Space and Matter*, New York
Cahill, R.T. / Klinger, C.M., 1996: Pregeometric Modelling of the Spacetime Phenomenology, *Physics Letters* **A 223**, 313-319; also: arXiv: gr-qc/9605018
Cahill, R.T. / Klinger, C.M., 1997: Bootstrap Universe from Self-Referential Noise, arXiv: gr-qc/9708013
Cahill, R.T. / Klinger, C.M., 1998: Self-Referential Noise and the Synthesis of Three-Dimensional Space, arXiv: gr-qc/9812083
Cahill, R.T. / Klinger, C.M., 2005: Bootstrap Universe from Self-Referential Noise, *Progress in Physics*, **2**, 108-112
Callender, C. / Huggett, N. (Eds.), 2001: *Physics meets Philosophy at the Planck Scale. Contemporary Theories of Quantum Gravity*, Cambridge
Callender, C. / Huggett, N., 2001a: Introduction, in: Callender / Huggett (2001)
Callender, C. / Huggett, N., 2001b: Why quantize Gravity (or any other field for the matter)?, *Philosophy of Science* (Proceedings), **68**, S382-S394
Cartwright, N., 1983: *How the Laws of Physics lie*, Oxford
Cartwright, N., 1989: *Nature's Capacities and Their Measurement*, Oxford
Cartwright, N., 1994: Fundamentalism vs the Patchwork of Laws, *Proceedings of the Aristotelian Society* **93**, 279-292
Cartwright, N., 1999: *The Dappled World*, Cambridge
DeWitt, B.S., 1967: Quantum Theory of Gravity. I. The Canonical Theory, *Physical Review* **160**, 1113-1148
DeWitt, B.S., 1967a: Quantum Theory of Gravity. II. The Manifestly Covariant Theory, *Physical Review* **162**, 1195-1239
DeWitt, B.S., 1967b: Quantum Theory of Gravity. III. Applications of the Covariant Theory, *Physical Review* **162**, 1239-1256
Dine, M., 2004: Is there a String Theory Landscape: Some Cautionary Remarks, arXiv: hep-th/0402101
Douglas, M.R., 2003: The Statistics of String/M Theory Vacua, *Journal for High Energy Physics* **0305**:046; also: arXiv: hep-th/0303194
Dreyer, O., 2004: Background Independent Quantum Field Theory and the Cosmological Constant Problem, arXiv: hep-th/0409048
Dreyer, O., 2006: Emergent General Relativity, arXiv: gr-qc/0604075
Dreyer, O., 2007: Why Things Fall, arXiv: 0710.4350 [gr-qc]
Earman, J., 1986: Why Space is not a Substance (at least not to the first degree), *Pacific Philosophical Quarterly* **67**, 225-244
Earman, J., 1989: *World Enough and Spacetime – Absolute Versus Relational Theories of Space and Time*, Cambridge, Ma.
Earman, J., 2002: Thoroughly Modern McTaggart. Or what McTaggart would have said if he had learned General Relativity Theory, *Philosopher's Imprint* **2**, 1-28, http://www.philosophersimprint.org
Earman, J., 2006: Two Challenges to the Requirement of Substantive General Covariance, *Synthese* **148**, 443-468

Earman, J., 2006a: The Implications of General Covariance for the Ontology and Ideology of Spacetime, in: D. Dieks (Ed.): *The Ontology of Spacetime*, Amsterdam, 3-23

Earman, J. / Norton, J.D., 1987: What Price Spacetime Substantivalism? – The Hole Story, *British Journal for the Philosophy of Science* **38**, 515-525

Ehlers, J. / Friedrich, H. (Eds.), 1994: *Canonical Gravity – From Classical to Quantum*, Berlin

Eling, C., 2008: Hydrodynamics of Spacetime and Vacuum Viscosity, arXiv: 0806.3165 [hep-th]

Eling, C. / Guedens, R. / Jacobson, T., 2006: Non-Equilibrium Thermodynamics of Spacetime, *Physical Review Letters* **96**, 121301; also: arXiv: gr-qc/0602001

Finkelstein, D.R., 1996: *Quantum Relativity: A Synthesis of the Ideas of Einstein and Heisenberg*, Berlin

Girelli, F. / Liberati, S. / Sindoni, L., 2008: On the Emergence of Time and Gravity, arXiv: 0806.4239 [gr-qc]

Green, M.B. / Schwarz, J.H. / Witten, E., 1987: *Superstring Theory*, 2 Vols., Cambridge

Hardy, L., 2007: Quantum Gravity Computers: On the Theory of Computation with Indefinite Causal Structure, arXiv: quant-ph/0701019

Hauser, A. / Corichi, A., 2005: Bibliography of Publications related to Classical Self-Dual Variables and Loop Quantum Gravity, arXiv: gr-qc/0509039

Hawkins, E. / Markopoulou, F. / Sahlmann, H., 2003: Evolution in Quantum Causal Histories, arXiv: hep-th/0302111

Hedrich, R., 2002: Anforderungen an eine physikalische Fundamentaltheorie, *Zeitschrift für Allgemeine Wissenschaftstheorie / Journal for General Philosophy of Science* **33/1**, 23-60

Hedrich, R., 2006: String Theory – From Physics to Metaphysics, *Physics and Philosophy* (Online Journal) 2006, 005; also: arXiv: physics/0604171

Hedrich, R., 2007: *Von der Physik zur Metaphysik – Physikalische Vereinheitlichung und Stringansatz*, Frankfurt am Main / Paris / Ebikon / Lancaster / New Brunswick

Hedrich, R., 2007a: The Internal and External Problems of String Theory – A Philosophical View, *Journal for General Philosophy of Science* **38,** 261-278; also: arXiv: physics/0610168

Henneaux, M. / Teitelboim, C., 1992: *Quantization of Gauge Systems*, Princeton

Henson, J., 2006: The Causal Set Approach to Quantum Gravity, arXiv: gr-qc/0601121

Hsu, S.D.H., 2007: Information, Information Processing and Gravity, arXiv: 0704.1154 [hep-th]

Hu, B.L., 2005: Can Spacetime be a Condensate, *International Journal of Theoretical Physics* **44**, 1785-1806; also: arXiv: gr-qc/0503067

Hu, B.L. / Verdaguer, E., 2003: Stochastic Gravity: A Primer with Applications, *Classical and Quantum Gravity* **20**, R1-R42; also: arXiv: gr-qc/0211090

Hu, B.L. / Verdaguer, E., 2004: Stochastic Gravity: Theory and Applications, *Living Reviews in Relativity* (Electronic Journal) **7/3**, www.livingreviews.org; also: arXiv: gr-qc/0307032

Hu, B.L. / Verdaguer, E., 2008: Stochastic Gravity: Theory and Applications, arXiv: 0802.0658 [gr-qc]

Isham, C.J., 1993: Canonical Quantum Gravity and the Problem of Time, in: *Integrable Systems, Quantum Groups, and Quantum Field Theory*, Dordrecht, 157-288; also: arXiv: gr-qc/9210011

Jacobson, E., 1995: Thermodynamics of Spacetime: The Einstein Equation of State, *Physical Review Letters* **75**, 1260-1263; also: arXiv: gr-qc/9504004

Jacobson, T., 1999: On the Nature of Black Hole Entropy, arXiv: gr-qc/9908031

Jacobson, T. / Parentani, R., 2003: Horizon Entropy, *Foundations of Physics* **33**, 323

Kaku, M., 1999: *Introduction to Superstrings and M-Theory*, 2nd Ed., New York

Kaplunovsky, V. / Weinstein, M., 1985: Space-Time: Arena or Illusion?, *Physical Review* **D 31**, 1879-1898

Kiefer, C., 2004: *Quantum Gravity*, Oxford ($^2$2007)

Kiefer, C., 2005: Quantum Gravity: General Introduction and Recent Developments, *Annalen der Physik* **15**, 129-148; also: arXiv: gr-qc/0508120

Konopka, T. / Markopoulou, F. / Severini, S., 2008: Quantum Graphity: A Model of Emergent Locality, *Physical Review* **D 77**, 104029; also: arXiv: 0801.0861 [hep-th]

Konopka, T. / Markopoulou, F. / Smolin, L., 2006: Quantum Graphity, arXiv: hep-th/0611197

Kribs, D.W. / Markopoulou, F., 2005: Geometry from Quantum Particles, arXiv: gr-qc/0510052

Kuchar, K., 1986: Canonical Geometrodynamics and General Covariance, *Foundations of Physics* **16**, 193-208

Kuchar, K., 1991: The Problem of Time in Canonical Quantization of Relativistic Systems, in: A. Ashtekar / J. Stachel (Eds.): *Conceptual Problems of Quantum Gravity*, Boston, 141-171

Kuchar, K., 1992: Time and Interpretation of Quantum Gravity, in: G. Kunstatter et al. (Eds.): *Proceedings of the 4th Canadian Conference on General Relativity and Relativistic Astrophysics*, Singapore

Kuchar, K., 1993: Canonical Quantum Gravity, arXiv: gr-qc/9304012

Livine, E.R. / Oriti, D., 2003: Implementing Causality in the Spin Foam Quantum Geometry, *Nuclear Physics* **B 663**, 231-279; also: arXiv: gr-qc/0210064

Livine, E.R. / Terno, D.R., 2007: Quantum Causal Histories in the Light of Quantum Information, *Physical Review* **D75**, 084001; also: arXiv: gr-qc/0611135

Lloyd, S., 1999: Universe as Quantum Computer, arXiv: quant-ph/9912088

Lloyd, S., 2005: A Theory of Quantum Gravity based on Quantum Computation, arXiv: quant-ph/0501135

Lloyd, S., 2007: *Programming the Universe: A Quantum Computer Scientist takes on the Cosmos*, New York

Markopoulou, F., 2000: The Internal Description of a Causal Set: What the Universe looks like from the inside, *Communications in Mathematical Physics* **211**, 559; also: arXiv: gr-qc/9811053

Markopoulou, F., 2000a: Quantum Causal Histories, *Classical and Quantum Gravity* **17**, 2059; also: arXiv: hep-th/9904009

Markopoulou, F., 2000b: An Insider's Guide to Quantum Causal Histories, *Nuclear Physics* **88** (Proc. Suppl.), 308-313; also: arXiv: hep-th/9912137

Markopoulou, F., 2004: Planck-Scale Models of the Universe, in: J.D. Barrow et al. (Eds.): *Science and Ultimate Reality: Quantum Theory, Cosmology, and Complexity*, Cambridge, 550-563; also: arXiv: gr-qc/0210086

Markopoulou, F., 2006: Towards Gravity from the Quantum, arXiv: hep-th/0604120

Markopoulou, F., 2007: New Directions in Background Independent Quantum Gravity, arXiv: gr-qc/0703097

Markopoulou, F. / Smolin, L., 1997: Causal Evolution of Spin Networks, *Nuclear Physics* **B 508**, 409; also: arXiv: gr-qc/9702025

Markopoulou, F. / Smolin, L., 1999: Holography in a Quantum Spacetime, arXiv: hep-th/9910146

Marolf, D., 2004: Resource Letter: The Nature and Status of String Theory, *American Journal of Physics* **72**, 730; also: arXiv: hep-th/0311044

Nicolai, H. / Peeters, K., 2006: Loop and Spin Foam Quantum Gravity, arXiv: hep-th/0601129

Nicolai, H. / Peeters, K. / Zamaklar, M., 2005: Loop Quantum Gravity: An Outside View, *Classical and Quantum Gravity* **22**, R193; also: arXiv: hep-th/0501114

Norton, J.D., 1988: The Hole Argument, in: A. Fine / J. Leplin (Eds.): *PSA 1988*, Vol. 2, East Lansing, 56-64

Norton, J.D., 1993: General Covariance and the Foundations of General Relativity, *Reports on Progress in Physics* **56**, 791-858

Norton, J.D., 2004: The Hole Argument, in: E.N. Zalta (Ed.): *Stanford Encyclopedia of Philosophy*, http://plato.stanford.edu

Oriti, D., 2001: Spacetime Geometry from Algebra: Spin Foam Models for non-perturbative Quantum Gravity, *Reports on Progress in Physics* **64**, 1489ff; also: arXiv: gr-qc/0106091

Oriti, D., 2003: *Spin Foam Models of Quantum Spacetime*, Ph.D. Thesis, arXiv: gr-qc/0311066

Padmanabhan, T., 2002: Gravity from Spacetime Thermodynamics, arXiv: gr-qc/0209088

Padmanabhan, T., 2004: Gravity as Elasticity of Spacetime: A Paradigm to understand Horizon Thermodynamics and Cosmological Constant, *International Journal of Modern Physics* **D 13**, 2293-2298; also: arXiv: gr-qc/0408051

Padmanabhan, T., 2007: Gravity as an Emergent Phenomenon: A Conceptual Description, arXiv: 0706.1654 [gr-qc]

Perez, A., 2003: Spin Foam Models for Quantum Gravity, *Classical and Quantum Gravity* **20**, R43-R104; also: arXiv: gr-qc/0301113

Perez, A., 2006: The Spin-Foam-Representation of Loop Quantum Gravity, arXiv: gr-qc/0601095

Peres, A. / Terno, D.R., 2001: Hybrid Classical-Quantum Dynamics, *Physical Review* **A 63**, 022101

Pesci, A., 2007: From Unruh Temperature to Generalized Bousso Bound, arXiv: 0708.3729 [gr-qc]

Pesci, A., 2008: On the Statistical-mechanical Meaning of Bousso Bound, arXiv: 0803.2642 [gr-qc]

Polchinski, J.G., 2000: *String Theory. Vol. 1: An Introduction to the Bosonic String*, Cambridge

Polchinski, J.G., 2000a: *String Theory. Vol. 2: Superstring Theory and Beyond*, Cambridge

Requardt, M., 1996: Discrete Mathematics and Physics on the Planck-Scale exemplified by means of a Class of 'Cellular Network Models' and their Dynamics, arXiv: hep-th/9605103

Requardt, M., 1996a: Emergence of Space-Time on the Planck-Scale Described as an Unfolding Phase Transition within the Scheme of Dynamical Cellular Networks and Random Graphs, arXiv: hep-th/9610055

Requardt, M., 2000: Let's Call it Nonlocal Quantum Physics, arXiv: gr-qc/0006063

Rickles, D., 2006: Time and Structure in Canonical Gravity, in: D. Rickles / S. French / J. Saatsi (Eds.): *The Structural Foundations of Quantum Gravity*, Oxford, 152-195

Rideout, D.P., 2002: Dynamics of Causal Sets, arXiv: gr-qc/0212064

Rideout, D.P. / Sorkin, R.D., 2000: A Classical Sequential Growth Dynamics for Causal Sets, *Physical Review* **D 61**, 024002; also: arXiv: gr-qc/9904062

Rideout, D.P. / Sorkin, R.D., 2001: Evidence for a Continuum Limit in Causal Set Dynamics, *Physical Review* **D 63**, 104011; also: arXiv: gr-qc/0003117

Rovelli, C., 1991: Time in Quantum Gravity: An Hypothesis, *Physical Review* **D 43**, 442-456

Rovelli, C., 1997: Loop Quantum Gravity, *Living Reviews in Relativity* (Electronic Journal), www.livingreviews.org; also: arXiv: gr-qc/9710008

Rovelli, C., 2002: Partial Observables, *Physical Review* **D 65**, 124013; also: arXiv: gr-qc/0110035

Rovelli, C., 2004: *Quantum Gravity*, Cambridge

Rovelli, C., 2007: Quantum Gravity, in: J. Earman / J. Butterfield (Eds.): *Handbook of the Philosophy of Science, Vol. 2: The Philosophy of Physics*, Amsterdam

Sakharov, A.D., 2000: Vacuum Quantum Fluctuations in Curved Space and the Theory of Gravitation, *General Relativity and Gravitation* **32**, 365-367 (Reprint; Original: *Doklady Akademii Nauk SSSR* **177** (1967) 70-71 / *Soviet Physics Doklady* **12** (1968) 1040-1041)

Sorkin, R.D., 2003: Causal Sets: Discrete Gravity, arXiv: gr-qc/0309009

Surya, S., 2007: Causal Set Topology, arXiv: 0712.1648 [gr-qc]

Susskind, L., 2004: Supersymmetry Breaking in the Anthropic Landscape, arXiv: hep-th/0405189

Susskind, L., 2005: *The Cosmic Landscape – String Theory and the Illusion of Intelligent Design*, New York

Tahim, M.O. et al., 2007: Spacetime as a deformable Solid, arXiv: 0705.4120 [gr-qc]

Terno, D.R., 2006: Inconsistency of Quantum-Classical Dynamics, and What it Implies, *Foundations of Physics* **36**, 102-111; also: arXiv: quant-ph/0402092

Thiemann, T., 2001: *Introduction to Modern Canonical Quantum General Relativity*, arXiv: gr-qc/0110034

Thiemann, T., 2002: Lectures on Loop Quantum Gravity, arXiv: gr-qc/0210094

Thiemann, T., 2006: Loop Quantum Gravity: An Inside View, arXiv: hep-th/0608210

Unruh, W. / Wald, R., 1989: Time and the Interpretation of Canonical Quantum Gravity, *Physical Review* **D 40**, 2598-2614

Visser, M., 2002: Sakharov's Induced Gravity: A Modern Perspective, *Modern Physics Letters* **A17**, 977-992; also: arXiv: gr-qc/0204062

Volovik, G.E., 2000: Links between Gravity and Dynamics of Quantum Liquids, arXiv: gr-qc/0004049

Volovik, G.E., 2001: Superfluid Analogies of Cosmological Phenomena, *Physics Reports* **352**, 195-348; also: arXiv: gr-qc/0005091

Volovik, G.E., 2003: *The Universe in a Helium Droplet*, Oxford

Volovik, G.E., 2006: From Quantum Hydrodynamics to Quantum Gravity, arXiv: gr-qc/0612134

Volovik, G.E., 2007: Fermi-Point Scenario for Emergent Gravity, arXiv: 0709.1258 [gr-qc]

Volovik, G.E., 2008: Emergent Physics: Fermi Point Scenario, arXiv: 0801.0724 [gr-qc]

Wald, R.M., 1994: *Quantum Field Theory on Curved Spacetime and Black Hole Thermodynamics*, Chicago

Wald, R.M., 2001: The Thermodynamics of Black Holes, *Living Reviews in Relativity* (Electronic Journal) **4/6**, www.livingreviews.org; also: arXiv: gr-qc/9912119

Weinberg, S., 1992: *Dreams of a Final Theory*, New York (dt.: *Der Traum von der Einheit des Universums*, München, 1993)

Wheeler, J.A., 1979: Frontiers of Time, in: N. Toraldo di Francia (Ed.): *Problems in the Foundations of Physics. Proceedings of the International School of Physics 'Enrico Fermi'*, Course 72, Amsterdam

Wheeler, J.A., 1983: Law without Law, in: J.A. Wheeler / W.H. Zurek (Eds.): *Quantum Theory and Measurement*, Princeton, N.J.

Wheeler, J.A., 1989: Information, Physics, Quantum: the Search for Links, in: *Proceedings 3rd International Symposium on the Foundation of Quantum Mechanics*, Tokyo, 354-368; also in: W.H. Zurek (Ed.): *Complexity, Entropy and the Physics of Information*, New York (1990), 3-28

Zhang, S.-C., 2002: To see a World in a Grain of Sand, arXiv: hep-th/0210162

Zizzi, P.A., 2001: Quantum Computation toward Quantum Gravity, *General Relativity and Gravitation* **33**, 1305-1318; also: arXiv: gr-qc/0008049

Zizzi, P.A., 2004: Computability at the Planck Scale, arXiv: gr-qc/0412076

Zizzi, P.A., 2005: A Minimal Model for Quantum Gravity, *Modern Physics Letters* **A 20**, 645-653; also: arXiv: gr-qc/0409069